\documentclass[%
 reprint,
superscriptaddress,
showpacs,
 amsmath,amssymb,
 aps,
]{revtex4-1}

\usepackage{epsfig}
\usepackage{epstopdf}
\usepackage{color}
\usepackage[normalem]{ulem}
\usepackage{dcolumn}
\usepackage{bm}

\newcommand\la{\langle}
\newcommand\ra{\rangle}
\newcommand\beq{\begin{eqnarray}}
\newcommand\eeq{\end{eqnarray}}
\newcommand\beqs{\begin{eqnarray*}}
\newcommand\eeqs{\end{eqnarray*}}

\def\Pslash{\rlap/{\mkern-1mu P}}

\def\Sslash{\rlap/{\mkern-1mu S}}

\def\nn{\nonumber}
\def\del{\partial}
\def\epal{\epsilon^{\alpha S_\perp w P_h}}

\def\sh{\hat{s}}
\def\th{\hat{t}}
\def\uh{\hat{u}}
\begin{document}

\preprint{APS/123-QED}

\title{Twist-3 fragmentation contribution to polarized hyperon production \\
in unpolarized hadronic collisions}

\author{Yuji Koike}
 \email{koike@phys.sc.niigata-u.ac.jp}
 \affiliation{Department of Physics, Niigata University, Ikarashi 2-no-cho, Nishi-ku, Niigata 950-2181, Japan}

\author{Andreas Metz}
 \email{metza@temple.edu}
 \affiliation{Department of Physics, Temple University, Philadelphia, PA, 19122, USA}

\author{Daniel Pitonyak}
 \email{dap67@psu.edu}
 \affiliation{Division of Science, Penn State University-Berks, Reading, PA 19610, USA}

\author{Kenta Yabe}
 \email{k.yabe.hadron@gmail.com}
 \affiliation{Graduate School of Science and Technology, Niigata University, Ikarashi 2-no-cho, Nishi-ku, Niigata 950-2181, Japan}

\author{Shinsuke Yoshida}
 \email{shinyoshida85@gmail.com}
  \affiliation{Theoretical Division, Los Alamos National Laboratory, Los Alamos, NM 87545, USA}

\date{\today}

\begin{abstract}
It has been known for a long time that hyperons produced in hadronic collisions are polarized 
perpendicular to the production plane of the reaction.
This effect cannot be described by using twist-2 collinear parton correlators only.
Here we compute the contribution of twist-3 fragmentation functions to the production of 
transversely polarized hyperons in unpolarized proton-proton collisions in the framework of 
collinear factorization. 
By taking into account the relations among the relevant twist-3 fragmentation functions 
which follow from the QCD equation of motion and the Lorentz invariance property of the correlators, 
we present the leading-order cross section for this term.
\end{abstract}



\maketitle

\section{Introduction} 
\label{s:introduction}
The first observation of transversely polarized hyperons in unpolarized hadronic collisions was 
already made in the 1970s.
Specifically, when colliding protons with a beryllium target and detecting $\Lambda$ hyperons, 
it was found that the $\Lambda$'s show a transverse polarization asymmetry (often denoted as $A_N$), 
which is largest (up to 30\%) for polarization perpendicular to the reaction plane and vanishes 
in the reaction plane~\cite{Bunce:1976yb}.
This pioneering measurement was followed by a number of corresponding experiments which, 
in particular, also covered different kinematic 
ranges~\cite{Heller:1978ty,Erhan:1979xm,Heller:1983ia,Lundberg:1989hw,
Yuldashev:1990az,Ramberg:1994tk,Fanti:1998px,Abt:2006da,Aaij:2013oxa,ATLAS:2014ona}. 
Some of the earlier data are reviewed in~\cite{Pondrom:1985aw,Panagiotou:1989sv}.
We also refer to~\cite{Metz:2016swz} for a list of relevant papers.
Generally $A_N$ vanishes for exact mid-rapidity of the hyperon in a process like 
$p \, p \to \Lambda^{\uparrow} X$, and it increases with increasing rapidity.
Hyperon polarization was also studied in related reactions such as 
$\gamma \, p \to \Lambda^{\uparrow} X$~\cite{Aston:1981em,Abe:1983jy}, 
quasi-real photo-production of $\Lambda$'s in 
lepton scattering off nucleons and nuclei~\cite{Airapetian:2007mx,Airapetian:2014tyc}, 
and in electron-positron collisions~\cite{Ackerstaff:1997nh,Abdesselam:2016nym}.

For high-energy collisions and sufficiently large transverse momentum $P_T$ of the hyperon, 
$A_N$ can be computed in perturbative quantum chromodynamics (QCD).
However, it has been known for a long time that by only using collinear 
leading-twist (twist-2) parton correlators one cannot describe this type of transverse 
single-spin asymmetry (SSA)~\cite{Kane:1978nd}.
As $A_N$ is a genuine twist-3 observable, one rather needs the full machinery of 
collinear higher-twist factorization~\cite{Efremov:1981sh,Ellis:1982wd,Ellis:1982cd}.
Already in the early 1980s this approach was used in connection with transverse 
SSAs~\cite{Efremov:1981sh,Efremov:1984ip}. 
Later works further elaborated on these twist-3 calculations, where a main focus was on 
the transverse target SSA for processes like $p^{\uparrow}p \to \pi X$ --- see for 
instance~\cite{Qiu:1991pp,Qiu:1991wg,Qiu:1998ia,Kanazawa:2000kp,Eguchi:2006qz,Eguchi:2006mc,
Kouvaris:2006zy,Koike:2009ge,Metz:2012ct,Beppu:2013uda}.
An overview of these calculations can be found in~\cite{Pitonyak:2016hqh}.

In collinear factorization, the transverse SSAs receive, {\it a priori}, twist-3 contributions from 
two-parton and three-parton correlation functions which are associated with 
either the initial-state or final-state hadrons. 
These correlators are parameterized in terms of twist-3 parton distribution functions (PDFs) 
and fragmentation functions (FFs), respectively.
While the complete leading-order (LO) twist-3 cross section for $p^{\uparrow}p \to \pi X$ 
can be found in the literature~\cite{Kouvaris:2006zy,Koike:2009ge,Metz:2012ct,Beppu:2013uda}, 
only part of the twist-3 cross section for $p \, p \to \Lambda^{\uparrow}X$ is 
available~\cite{Kanazawa:2000cx,Zhou:2008fb,Koike:2015zya}.
The present work is a major step towards
completing the calculation of all possible terms.

The numerator of the transverse SSA for $p \, p \to \Lambda^{\uparrow}X$ has two types of contributions.
The first one, which contains a twist-3 PDF for one of the unpolarized protons 
combined with the unpolarized twist-2 PDF for the other proton and 
the spin-dependent twist-2 ``transversity''
FF, was derived in Refs.~\cite{Kanazawa:2000cx,Zhou:2008fb,Koike:2015zya}.
Here we focus on the second contribution, which involves twist-3 FFs and the twist-2 unpolarized PDFs of 
the protons.  
(A short version of the present
work was presented in \cite{Yabe:talk,Yabe:proceedings}.)
Specifically, we compute all the LO terms that are related to quark-gluon-quark ({\it qgq}) 
fragmentation correlators, while terms given by quark-antiquark-gluon ($q\bar{q}g$) correlators 
and pure gluon ({\it gg} and {\it ggg}) correlators will be considered elsewhere.

Not only is our study important for obtaining a complete analytical result, but it 
may also be critical for the phenomenology of this observable.
In this context we emphasize that recent work strongly suggests numerical dominance of the 
collinear twist-3 fragmentation contribution for $A_N$ in $p^{\uparrow}p \to \pi X$~\cite{
Kanazawa:2014dca,Gamberg:2017gle}. 

The remainder of the paper is organized as follows:  
In Sec.~\ref{s:definition}, we list the definitions of the twist-3 FFs that are relevant for the present study. 
In that section we also present relations among the FFs which are based on the QCD equation of motion 
and Lorentz invariance~\cite{Kanazawa:2015ajw}. 
These relations are crucial for, in particular, obtaining a frame-independent result for $A_N$.
In Sec.~\ref{s:calculation}, we discuss the calculation for the twist-3 fragmentation contribution to the 
cross section for $p \, p\to \Lambda^{\uparrow}X$, while Sec.~\ref{s:summary} is devoted to a brief summary.

\section{Twist-3 fragmentation functions and their relations}
\label{s:definition}
We first recall the definitions of the twist-3 FFs for a transversely polarized spin-$\frac{1}{2}$ hadron.
One can identify, {\it a priori}, three different types of such functions, 
which in~\cite{Kanazawa:2015ajw} were referred to as {\it intrinsic}, {\it kinematical}, and {\it dynamical} FFs.
We start with the intrinsic twist-3 FFs for $\Lambda^{\uparrow}$.
They are defined through a quark-quark ({\it qq}) fragmentation 
correlator according to~\cite{Mulders:1995dh,Bacchetta:2006tn,Kanazawa:2015ajw} 
\begin{eqnarray}
\Delta_{ij}(z)&=&{1\over N}\sum_X\!\int \!\!\frac{d\lambda}{2\pi} e^{-i{\lambda\over z}}\la 0|\psi_i(0)| 
h(P_h,S_\perp)X\ra\nn\\
&&\hspace{0.1cm}\times\la h(P_h,S_\perp)X|\bar{\psi}_j(\lambda w)| 0\ra\nn\\[0.3cm]
&=&\left(\gamma_5\Sslash_\perp\frac{\Pslash_h}{z}\right)_{\!\!ij}\!\!H_1(z)+M_h
\epal(\gamma_\alpha)_{ij}\frac{D_T(z)}{z}\nn\\
&&\hspace{0.1cm}+\,M_h(\gamma_5\Sslash_\perp)_{ij}\frac{G_T(z)}{z}+\cdots,
\label{e:ff1}
\end{eqnarray}
where $\psi_i$, $\psi_j$ are the quark fields carrying the spinor indices $i$, $j$, and a
color average is implied in Eq.~(\ref{e:ff1}), with $N = 3$ the number of quark colors.
The hadron ($\Lambda$) is characterized by its four-momentum $P_h$ 
and (transverse) spin vector $S_\perp$, while $M_h$ is its mass.
The four-vector $w^{\mu}$ is light-like ($w^2=0$) and satisfies $P_h \cdot w=1$.
For simplicity, Wilson lines in the operator (\ref{e:ff1}) are suppressed.
Here and below we use the shorthand notation 
$\epal\equiv\epsilon^{\alpha\beta\gamma\delta}S_{\perp\beta}w_\gamma P_{h\delta}$, where our
convention for the Levi-Civita tensor is $\epsilon^{0123} = +1$.
The r.h.s.~of~Eq.~(\ref{e:ff1}) contains the twist-2 transversity FF $H_1$, 
which describes the probability for a transversely polarized quark to fragment into a transversely 
polarized hadron, and the intrinsic twist-3 FFs $D_T$ and $G_T$.
These (dimensionless) FFs depend on the fraction $z$ of the quark momentum which is carried by the hadron.
Hermiticity implies that they are real-valued.
From the functions available in (\ref{e:ff1}), it is actually only the na\"ive 
time-reversal-odd (T-odd) function $D_T$ which contributes to the piece of the 
transverse SSA we calculate here.  

The kinematical twist-3 FFs parameterize the derivative of 
the {\it qq} correlator~\cite{Mulders:1995dh,Bacchetta:2006tn,Kanazawa:2015ajw}, 
\begin{eqnarray}
&&\Delta_{\del ij}^\alpha(z)
={1\over N}\sum_X\!\!\int \!\!\frac{d\lambda}{2\pi} 
e^{-i{\lambda\over z}}\la 0|[\infty w,0]\psi_i(0)| h(P_h,S_\perp)X \ra \nn\\
&&\qquad\quad\quad\quad\times\la h(P_h,S_\perp)X|\bar{\psi}_j(\lambda w)
[\lambda w,\infty w]| 0\ra\overleftarrow{\del}^\alpha\nn\\
&&\qquad\qquad =-iM_h\epal(\Pslash_h)_{ij}\frac{D_{1T}^{\perp(1)}(z)}{z}\nonumber\\
&&\qquad\quad\quad\quad +\, iM_hS_\perp^\alpha(\gamma_5\Pslash_h)_{ij}\frac{G_{1T}^{\perp(1)}(z)}{z}+\cdots. 
\label{e:ff2}
\end{eqnarray}
The derivative on the r.h.s.~of~(\ref{e:ff2}) also acts on the Wilson line, which generally is defined through
\begin{equation}
[0,\lambda\omega]=\mathcal{P} \, {\rm exp}\left\{i{\rm g}\!\int^0_\lambda\!\!dt \, \omega_\mu A^\mu(t\omega)\right\} \,,
\end{equation}
where $\mathcal{P}$ indicates path-ordering and $g$ is the strong coupling.
The FFs $D_{1T}^{\perp(1)}$ and $G_{1T}^{\perp(1)}$ are also real-valued.
This (T-odd) function is a particular moment of a transverse-momentum-dependent (TMD) FF~\cite{Mulders:1995dh,Bacchetta:2006tn},
\begin{equation}
D_{1T}^{\perp(1)}(z)=z^2 \int d^2\vec{p}_\perp \frac{{\vec{p}_\perp}^{\;2}}{2M_h^2}
D_{1T}^\perp(z,z^2\vec{p}_\perp^{\;2}) \,,
\end{equation}
with $D_{1T}^\perp$ describing the fragmentation of an unpolarized quark 
into a transversely polarized spin-$\frac{1}{2}$ hadron. 
Using the so-called generalized parton model, which exclusively works with TMD PDFs and 
FFs, in Ref.~\cite{Anselmino:2000vs} this function was fitted to $A_N$ data for $p \, p \to \Lambda^{\uparrow} X$.
The result of the fit was then used to estimate transverse SSAs in semi-inclusive 
deep-inelastic scattering~\cite{Anselmino:2001js}, including neutrino-nucleon scattering 
$\nu N \to \ell^{\pm} \Lambda^{\uparrow} X$ for which some data are available~\cite{Astier:2000ax}.
We note that, like with the intrinsic functions, 
only T-odd kinematical correlators can contribute to our calculation.  
Namely, only $D_{1T}^{\perp(1)}$ can enter from (\ref{e:ff2}). 

Let us finally discuss the dynamical twist-3 FFs.
They parameterize the so-called $F$-type {\it qgq} correlator 
(see, for instance, Refs.~\cite{Meissner:thesis,Kanazawa:2015jxa,Kanazawa:2015ajw}), 
\begin{eqnarray}
&&\hspace{-0.4cm}\Delta_{F ij}^\alpha(z,z_1)\nn\\
&=&{1\over N}\sum_X\!\int \!\!\frac{d\lambda}{2\pi}\! \int \!\!\frac{d\mu}{2\pi} 
e^{-i{\lambda\over z_1}}e^{-i\mu({1\over z}-{1\over z_1})}\la 0|\psi_i(0)| h(P_h, S_\perp)X\ra\nn\\
&&\hspace{0.075cm}\times\,\la h(P_h,S_\perp)X|\bar{\psi}_j(\lambda w){\rm g}F^{\alpha w}(\mu w)| 0\ra\nn\\[0.3cm]
&=&M_h \epal(\Pslash_h)_{ij}\frac{\widehat{D}_{FT}^\ast(z,z_1)}{z}\nn\\
&&\hspace{0.075cm}-\,iM_h 
S_\perp^\alpha(\gamma_5\Pslash_h)_{ij}\frac{\widehat{G}_{FT}^\ast(z,z_1)}{z}+\cdots,
\label{e:ff3}
\end{eqnarray}
where $F^{\alpha w} \equiv F^{\alpha \beta} w_{\beta}$ represents a component of the gluon field strength tensor. 
The (two-argument) FFs $\widehat{D}_{FT}(z,z_1)$ and $\widehat{G}_{FT}(z,z_1)$ have support for 
$1>z>0$ and $z_1>z$~\cite{Gamberg:2008yt,Meissner:2008yf,Gamberg:2010uw,Kanazawa:2013uia}.   
These support properties imply, in particular, that the functions 
themselves~\cite{Meissner:2008yf} and their $z_1$-partial 
derivatives~\cite{Kanazawa:2015ajw} 
vanish for a vanishing gluon momentum.
Therefore, for fragmentation one has no so-called gluonic 
poles~\cite{Qiu:1991pp,Qiu:1991wg}, which play a very important 
role for contributions to SSAs caused by twist-3 effects coming from initial-state hadrons.
The vanishing of gluonic poles in fragmentation is also intimately connected with the universality of 
TMD FFs (see, for instance, 
Refs.~\cite{Metz:2002iz,Collins:2004nx,Gamberg:2008yt,Meissner:2008yf,Gamberg:2010uw}).
In general, the dynamical twist-3 FFs are 
complex~\cite{Meissner:thesis,Yuan:2009dw,Kang:2010zzb,Metz:2012ct,Kanazawa:2013uia}, 
and we use their complex conjugate in Eq.~(\ref{e:ff3}).  
(In this paper we follow the convention of
\cite{Kanazawa:2015ajw} for $\widehat{D}_{FT}$ and $\widehat{G}_{FT}$.)
One can also define D-type twist-3 FFs by replacing $gF^{\alpha w}(\mu w)$ in~(\ref{e:ff3}) with the 
covariant derivative $D^{\alpha}(\mu w)$. 
These functions, however, do not represent new independent objects, but they can rather be related 
to the F-type functions.
For the imaginary parts of the FFs, which matter in the present study, one has the 
relations~\cite{Eguchi:2006qz,Eguchi:2006mc,Meissner:thesis,
Yuan:2009dw,Kang:2010zzb,Metz:2012ct,Kanazawa:2013uia}
\begin{eqnarray}
{\rm Im} \, \widehat{D}_{DT}(z,z_1) & = & P \frac{1}{1/z - 1/z_1} {\rm Im} \, \widehat{D}_{FT}(z,z_1) 
\nonumber \\
& & - \; \delta\Big( \frac{1}{z} - \frac{1}{z_1} \Big) D_{1T}^{\perp (1)}(z) \,,
\label{e:DD_DF} \\ 
{\rm Im} \, \widehat{G}_{DT}(z,z_1) & = & P \frac{1}{1/z - 1/z_1} {\rm Im} \, \widehat{G}_{FT}(z,z_1) \,,
\label{e:GD_GF}
\end{eqnarray}
where here $P$ indicates the principal-value prescription.
For the calculation in Sec.~\ref{s:calculation} we will use the F-type FFs. 

Although, {\it a priori}, all three types of FFs as defined in (\ref{e:ff1}), (\ref{e:ff2}), 
(\ref{e:ff3}) appear in the derivation of the twist-3 cross section for $p \, p\to\Lambda^{\uparrow}X$, 
these functions are not independent.
There exist relations based on the QCD equation of motion (e.o.m.) and so-called Lorentz invariance relations (LIRs).
A comprehensive derivation of these relations as well as a list of references can be found in~\cite{Kanazawa:2015ajw}.
The e.o.m.~relation which is relevant for the present study takes its simplest form for the D-type functions, 
\begin{equation}
\int\frac{dz_1}{z_1^2} \Bigl({\rm Im} \, \widehat{D}_{DT}(z,z_1)-{\rm Im} \, \widehat{G}_{DT}(z,z_1)\Bigr)
=  \frac{D_T(z)}{z} \,.
\label{e:eom_D}
\end{equation}
Using Eqs.~(\ref{e:DD_DF}), (\ref{e:GD_GF}) one finds the e.o.m.~relation involving the F-type functions,
\begin{eqnarray}
\int_z^\infty\frac{dz_1}{z_1^2}\frac{1}{1/z-1/z_1}\Bigl({\rm Im} \, \widehat{D}_{FT}&&(z,z_1)-{\rm Im} \, 
\widehat{G}_{FT}(z,z_1)\Bigr)\nn\\
=\frac{D_T(z)}{z}+D_{1T}^{\perp(1)}(z) \,. 
\label{e:eom_F}
\end{eqnarray}
Making use of Lorentz invariance of the parton correlation functions one can further derive the LIR
\begin{eqnarray}
-{2\over z}\!\int_z^\infty\!\frac{dz_1}{z_1^2}\frac{{\rm Im}\widehat{D}_{FT}(z,z_1)}{(1/z_1-1/z)^2} 
=\frac{D_T(z)}{z}+\frac{d\!\left(\!D_{1T}^{\perp(1)}(z)/z\!\right)}{d(1/z)}.\nn\\  
\label{e:LIR}
\end{eqnarray}
These relations generally simplify the form of twist-3 cross sections and, in particular, guarantee their 
frame-independence~\cite{Kanazawa:2014tda,Kanazawa:2015jxa,Kanazawa:2015ajw,Gamberg:2017gle}.

\section{Twist-3 cross section for $pp\to\Lambda^\uparrow X$}
\label{s:calculation}
\begin{figure}[t]
\begin{center}
\vspace{0.8cm}

\includegraphics[width=7cm]{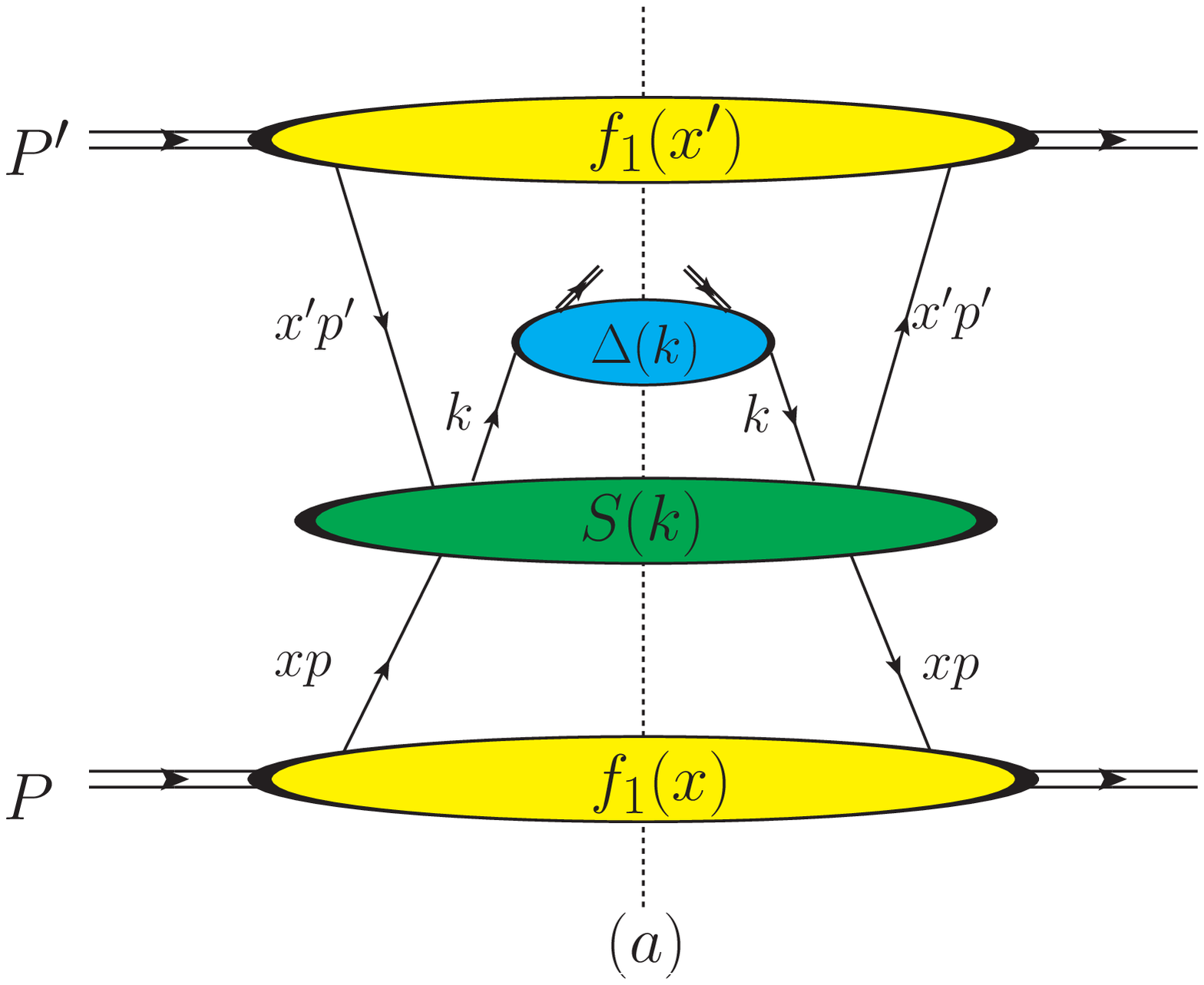}

\vspace{0.5cm}

\includegraphics[width=7cm]{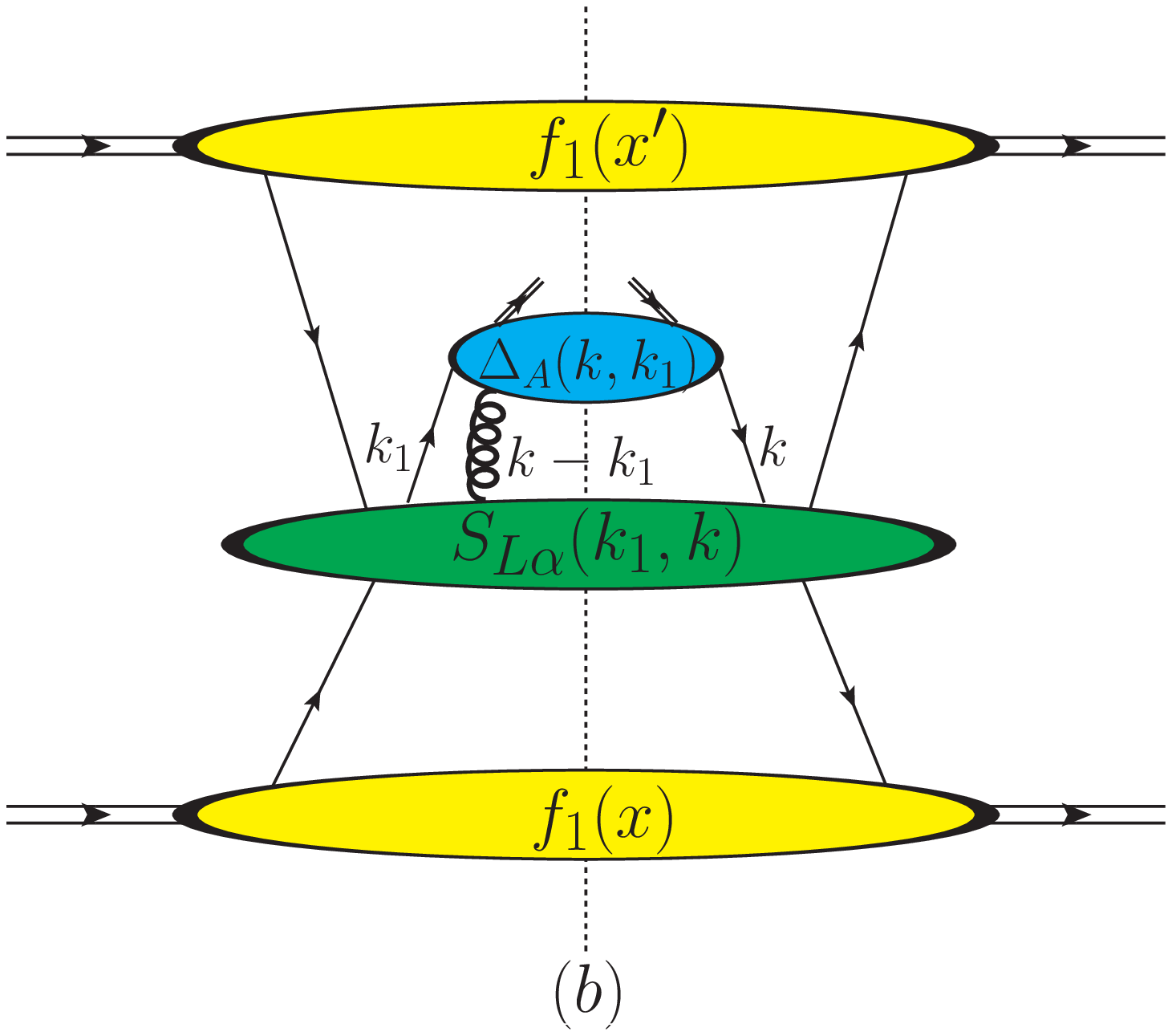}
\end{center}
 \caption{
Generic diagrams giving rise to the 
twist-3 fragmentation contribution to the 
polarized cross section for the process in~(\ref{e:ppLX}). 
The top blob and bottom blob indicate the unpolarized twist-2 PDFs in the protons. 
The second blob from the top represents the fragmentation matrix elements 
for $\Lambda$:~$\Delta(k)$ in (a) and $\Delta_A(k,k_1)$ in (b).   
The second blob from the bottom is the partonic hard scattering parts:~$S(k)$ 
in (a) and $S_{L\alpha}(k_1,k)$ in (b). 
The mirror diagram of $(b)$ also contributes and is included in the third term of (\ref{e:form1}).}
\label{f:t3ff}
\end{figure} 

We now sketch the derivation and present the results of the twist-3 spin-dependent cross section for
\begin{equation}
p(p) + p(p') \to \Lambda^{\uparrow}(P_h,S_\perp ) + X \,.
\label{e:ppLX}
\end{equation}
Applying the Feynman gauge formalism developed in~\cite{Kanazawa:2013uia}, 
one can obtain the corresponding cross section as
\begin{widetext}
\begin{eqnarray}
{P_h^0}\frac{d\sigma(P_h,S_{\perp})}{d^3P_h}&=&\frac{1}{16\pi^2s}\!\int\!\!\frac{dx}{x}f_1(x)\!\int\frac{dx'}{x'}f_1(x')
\biggl[\int\frac{dz}{z^2}{\rm Tr}\left[\Delta(z)S(P_h/z)\right]
-i\int\frac{dz}{z^2}{\rm Tr}\biggl[\Omega^{\alpha}_{\beta}\Delta^{\beta}_{\partial}(z)
\left.\frac{\partial S(k)}{\partial k^\alpha}\right|_{k=P_h/z}\biggr]\nn\\
&+& 2\,{\rm Re}\left\{(-i)\int\frac{dz}{z^2}\frac{dz_1}{z_1^2}{\rm Tr}\left[
\Omega^{\alpha}_{\beta}\Delta^{\beta}_{F}(z,z_1)P\!\left(\frac{1}{1/z_1-1/z}\right)
S_{L\alpha}\!\!\left(\frac{P_h}{z_1},\frac{P_h}{z}\right)\right]\right\}\biggr],
\label{e:form1}
\end{eqnarray}
\end{widetext}
where the summation over all channels and parton types is implicit.
In Eq.~(\ref{e:form1}), $s=(p+p')^2$ is the square of the center of 
mass energy and $\Omega^{\alpha}_{\beta}=g^{\alpha}_{\beta}-P_h^\alpha w_\beta$.
The unpolarized twist-2 PDF is denoted by $f_1$, and the correlators 
$\Delta(z)$, $\Delta^{\beta}_{\partial}(z)$ and $\Delta^{\beta}_{F}(z,z_1)$ 
are defined in (\ref{e:ff1}), (\ref{e:ff2}) and (\ref{e:ff3}).  
The symbol ${\rm Tr}$ indicates the trace over color and spinor indices.  
In deriving Eq.~(\ref{e:form1}), 
we introduced the partonic hard scattering parts (before collinear expansion),
$S(k)$ and $S_{L\alpha}\!\!\left(k_1,k\right)$, 
corresponding, respectively, 
to the fragmentation matrix elements
\begin{eqnarray}
\Delta_{ij}(k)&=&{1\over N} \sum_X \int\,d^4\xi\,e^{-ik\cdot\xi} \la 0| \psi_i(0)|h(P_h,S_\perp) X\ra
 \nonumber\\
&&\hspace{0.1cm}\times\,\la h(P_h,S_\perp) X|\bar{\psi}_j(\xi)|0\ra, \nonumber\\
\end{eqnarray}
and 
\begin{eqnarray}
&&\Delta_{A,ij}(k,k_1)={1\over N} \sum_X \int\,d^4\xi\int\,d^4\,\eta \,e^{-ik_1
\cdot\xi}\,e^{-i(k-k_1)\cdot\eta}\nonumber\\
&&\quad\times\la 0| \psi_i(0)|h(P_h,S_\perp) X\ra
\la h(P_h,S_\perp) X|\bar{\psi}_j(\xi)gA^\alpha (\eta) |0\ra, \nn\\
\end{eqnarray}
as shown in
Figs.~\ref{f:t3ff}~(a) and (b).  
In Ref.~\cite{Kanazawa:2013uia}, it has been proven that, after the collinear expansion,  
$S$ and $S_{L\alpha}$ eventually constitute
the partonic hard cross section for the gauge-invariant correlation functions 
$\Delta(z)$, $\Delta^{\beta}_{\partial}(z)$ and $\Delta^{\beta}_{F}(z,z_1)$, as shown in (\ref{e:form1}).  
Note $S_{L\alpha}\!\!\left(\frac{P_h}{z_1},\frac{P_h}{z}\right)$ is the hard part for the 
diagram in which the coherent gluon line from $\Delta^{\beta}_{F}(z,z_1)$ is located to 
the left of the cut, and the effect of the mirror diagram is taken into account by 
the principal value prescription and 
the factor 2 for the third term in (\ref{e:form1}).
Substituting (\ref{e:ff1}), (\ref{e:ff2}) and (\ref{e:ff3}) into Eq.~(\ref{e:form1}), 
one can cast the cross section in the following form:
\begin{widetext}
\begin{eqnarray}
{P_h^0}&&\frac{d\sigma(P_h,S_{\perp})}{d^3P_h}=\frac{\alpha_s^2M_h}{s}
\sum_{i=1}^2A^{(i)}(w)
\!\int\!\!\frac{dx}{x}f_1(x)\!\int\frac{dx'}{x'}f_1(x')\!\int\!\!\frac{dz}{z^3}\delta(\hat{s}+\hat{t}+\hat{u})\nn\\
&&\times\biggl[\frac{D_T(z)}{z}\hat{\sigma}_T^{(i)}-\frac{d}{d(1/z)}\frac{D_{1T}^{\perp(1)}(z)}{z}
\hat{\sigma}_D^{(i)}-D_{1T}^{\perp(1)}(z)\hat{\sigma}_{ND}^{(i)}
+\int_z^\infty\frac{dz_1}{z_1^2}\left(\frac{1}{1/z-1/z_1}\right)
{\rm Im}\widehat{D}_{FT}(z,z_1)\hat{\sigma}_{DF1}^{(i)}\nn\\
&&+\int_z^\infty\frac{dz_1}{z_1^2}\left(\frac{z_1}{1/z-1/z_1}\right){\rm Im}\widehat{D}_{FT}(z,z_1)\hat{\sigma}_{DSFP}^{(i)}
-\frac{2}{z}\int_z^\infty\frac{dz_1}{z_1^2}\left(\frac{1}{(1/z_1-1/z)^2}\right){\rm Im}\widehat{D}_{FT}(z,z_1)
\hat{\sigma}_{DF2}^{(i)}
\nn\\
&&+\int_z^\infty\frac{dz_1}{z_1^2}\left(\frac{1}{1/z-1/z_1}\right){\rm Im}\widehat{G}_{FT}(z,z_1)
\hat{\sigma}_{GF1}^{(i)}
+\int_z^\infty\frac{dz_1}{z_1^2}\left(\frac{z_1}{1/z-1/z_1}\right){\rm Im}\widehat{G}_{FT}(z,z_1)
\hat{\sigma}_{GSFP}^{(i)}\nn\\
&&-\frac{2}{z}\int_z^\infty\frac{dz_1}{z_1^2}\left(\frac{1}{(1/z_1-1/z)^2}\right)
{\rm Im}\widehat{G}_{FT}(z,z_1)\hat{\sigma}_{GF2}^{(i)}
\biggl] , 
\label{e:Xsec1}
\end{eqnarray}
\end{widetext}
where $A^{(1)}(w)\equiv{p'\cdot P_h\over p\cdot p'} \epsilon^{P_hpwS_\perp}$, 
$A^{(2)}(w)\equiv{p\cdot P_h\over p\cdot p'}\epsilon^{P_hwp'S_\perp}$, and each partonic 
cross section $\hat{\sigma}^{(i)}_Y$ ($i=1,\,2$, $Y=T,\,D,\,ND,\,\dots$) is a 
function of the partonic Mandelstam variables defined as $\hat{s}=(xp+x'p')^2$, 
$\hat{t}=(xp-P_h/z)^2$, $\hat{u}=(x'p'-P_h/z)^2$.  
The lowest-order Feynman diagrams for the partonic hard parts $S$ and $S_{L\alpha}$ 
in each channel are shown in Figs.~\ref{f:qq_qq_1}--\ref{f:qg_qg}.
Several comments are in order here:
\noindent
(i) Unlike in the case of the twist-3 PDFs, for twist-3 FFs the nonpole term of the hard 
scattering coefficients contributes to $A_N$.
(In this context, see also the discussion in the paragraph after Eq.~(\ref{e:ff3}).)
In particular, the imaginary part of the complex functions $\widehat{D}_{FT}$ and $\widehat{G}_{FT}$ 
contributes to the spin-dependent cross section, reflecting the na\"ive T-odd nature of $A_N$.  

\noindent
(ii) One finds that the $z_1$-dependence of the hard parts $S_{L\alpha}$ in Eq.~(\ref{e:form1}) 
has a relatively simple structure that does not ``mix'' with the partonic Mandelstam variables.
Therefore, the contributions from ${\rm Im} \, \widehat{D}_{FT}$ 
and ${\rm Im} \, \widehat{G}_{FT}$ can be brought into the form shown in (\ref{e:Xsec1}),
where the hard partonic cross sections $\hat{\sigma}^{(i)}_Y$ are independent of $z_1$ and
only depend on $\hat{s}$, $\hat{t}$, $\hat{u}$.

\noindent
(iii) Calculation of the diagrams provides the relations
$\hat{\sigma}^{(i)}_{DF1}=-\hat{\sigma}_{GF1}^{(i)}$, 
$\hat{\sigma}_{DSFP}^{(i)}=\hat{\sigma}_{GSFP}^{(i)}$, 
$\hat{\sigma}_{GF2}^{(i)}=0$ ($i=1,\,2$) for all channels.  
Using these relations in combination with the e.o.m.~relation (\ref{e:eom_F}) and the LIR (\ref{e:LIR})
one can rewrite the cross section in~(\ref{e:Xsec1}) in a very compact manner --- see Eq.~(\ref{e:Xsec2}) below ---
by introducing the combinations
$\hat{\sigma}_1^{(i)} \equiv\hat{\sigma}_T^{(i)}+\hat{\sigma}_{DF1}^{(i)}+\hat{\sigma}_{DF2}^{(i)}$,
$\hat{\sigma}_2^{(i)}\equiv\hat{\sigma}_D^{(i)}-\hat{\sigma}_{DF2}^{(i)}$,
$\hat{\sigma}_3^{(i)}\equiv\hat{\sigma}_{ND}^{(i)}-\hat{\sigma}_{DF1}^{(i)}$ and 
$\hat{\sigma}_4^{(i)}\equiv\hat{\sigma}_{DSFP}^{(i)}$.

\noindent
(iv) In order to test the frame-independence of our result, we have computed the cross section in two different frames:~$w_1^{\mu}=\frac{p'^{\mu}}{p'\cdot P_h}$ 
and $w_2^{\mu}=\frac{p^{\mu}}{p\cdot P_h}$.  In the first frame, $A^{(1)}(w_1)$ is nonzero while $A^{(2)}(w_1)$
vanishes, so the result depends only on $A^{(1)}(w_1)$ and $\hat{\sigma}_{1,2,3,4}^{(1)}(w_1)$.  In the second frame, $A^{(2)}(w_2)$ is nonzero while $A^{(1)}(w_2)$
vanishes, so the result depends only on $A^{(2)}(w_2)$ and $\hat{\sigma}_{1,2,3,4}^{(2)}(w_2)$.  Since $A^{(1)}(w_1) = A^{(2)}(w_2)$,  
and we also found $\hat{\sigma}_{1,2,3,4}^{(1)}(w_1)
=\hat{\sigma}_{1,2,3,4}^{(2)}(w_2)\equiv \hat{\sigma}_{1,2,3,4}$ for all channels, the result is the same in both frames.

Our final expression for the frame-independent twist-3 cross section reads 
\begin{eqnarray}
\hspace{-0.5cm}
&&P_h^0 \frac{d\sigma(P_h,S_{\perp})}{d^3P_h}=\frac{2\alpha_s^2M_h}{s^2}
\epsilon^{P_hpp'S_\perp} 
\!\int\!\!\frac{dx}{x}f_1(x)\!\int\frac{dx'}{x'}f_1(x')\nonumber\\
&&\times\,\int\!\!
\frac{dz}{z^3}
\delta(\hat{s}+\hat{t}+\hat{u})
\left[
\frac{D_T(z)}{z}
\hat{\sigma}_1
-\left\{\frac{d}{d(1/z)}\frac{D_{1T}^{\perp(1)}(z)}{z}\right\}
\hat{\sigma}_2\right.\nonumber\\
&&\left.\quad\quad
-\,D_{1T}^{\perp(1)}(z)
\hat{\sigma}_3
+\int_z^\infty\frac{dz_1}{z_1^2}\left(\frac{z_1}{1/z-1/z_1}\right)\right.\nn\\
&&\left.\qquad\qquad\times\left({\rm Im}\widehat{D}_{FT}(z,z_1)+{\rm Im}\widehat{G}_{FT}(z,z_1)\right)
\hat{\sigma}_4
\right].
\label{e:Xsec2}
\end{eqnarray}
This represents the complete result of the cross section caused by twist-3 effects 
of the {\it qq} and {\it qgq} fragmentation correlators depicted in Fig.~\ref{f:t3ff}.
The partonic cross sections for each channel are given as follows:
\begin{figure}[t]
\begin{center}
\includegraphics[width=6.0cm]{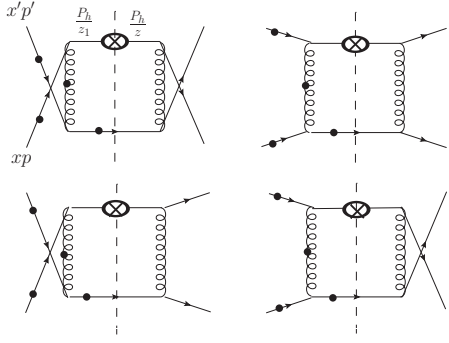}
\end{center}
\vspace{-0.5cm}
\caption{Feynman diagrams for the hard parts $S(k)$ and $S_{L\alpha}(k_1,k)$ in 
the $qq\to qq$ (all graphs), $qq'\to qq'$ (top left graph) and 
$qq'\to q'q$ (top right graph) channels. 
The circled cross indicates the fragmentation insertion. 
When ignoring the dots the diagrams determine the hard parts $S(k)$.
Graphs for $S_{L\alpha}(k_1,k)$ are obtained by attaching a 
coherent gluon line from the fragmentation function to one of the four 
dots in each diagram.}
\label{f:qq_qq_1}
\end{figure} 
\begin{figure}[!]
\begin{center}
\includegraphics[width=6cm]{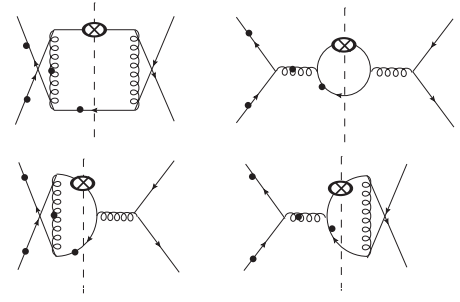}
\end{center}
\vspace{-0.5cm}
\caption{Same as Fig.~\ref{f:qq_qq_1}, but for the $q\bar{q}\to q\bar{q}$ (all graphs), 
$q\bar{q}'\to q\bar{q}'$ (top left graph) and $q\bar{q}\to q'\bar{q}'$ (top right graph) channels.}
\label{f:qq_qq_2}
\end{figure} 
\begin{figure}[!]
\begin{center}
\includegraphics[width=8cm]{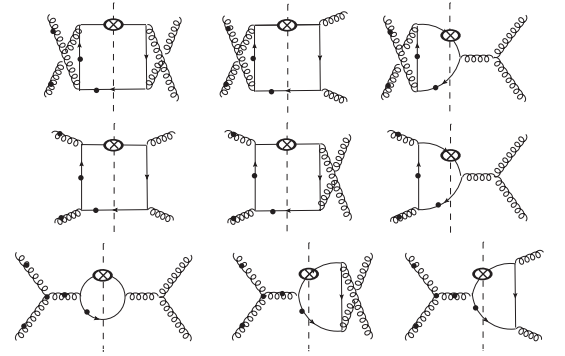}\hspace{1cm}
\end{center}
\vspace{-0.5cm}
\caption{Same as Fig.~\ref{f:qq_qq_1}, but for the $gg\to q\bar{q}$ channel.}
\label{f:gg_qq}
\end{figure} 
\begin{figure}[!]
\begin{center}
\includegraphics[width=8cm]{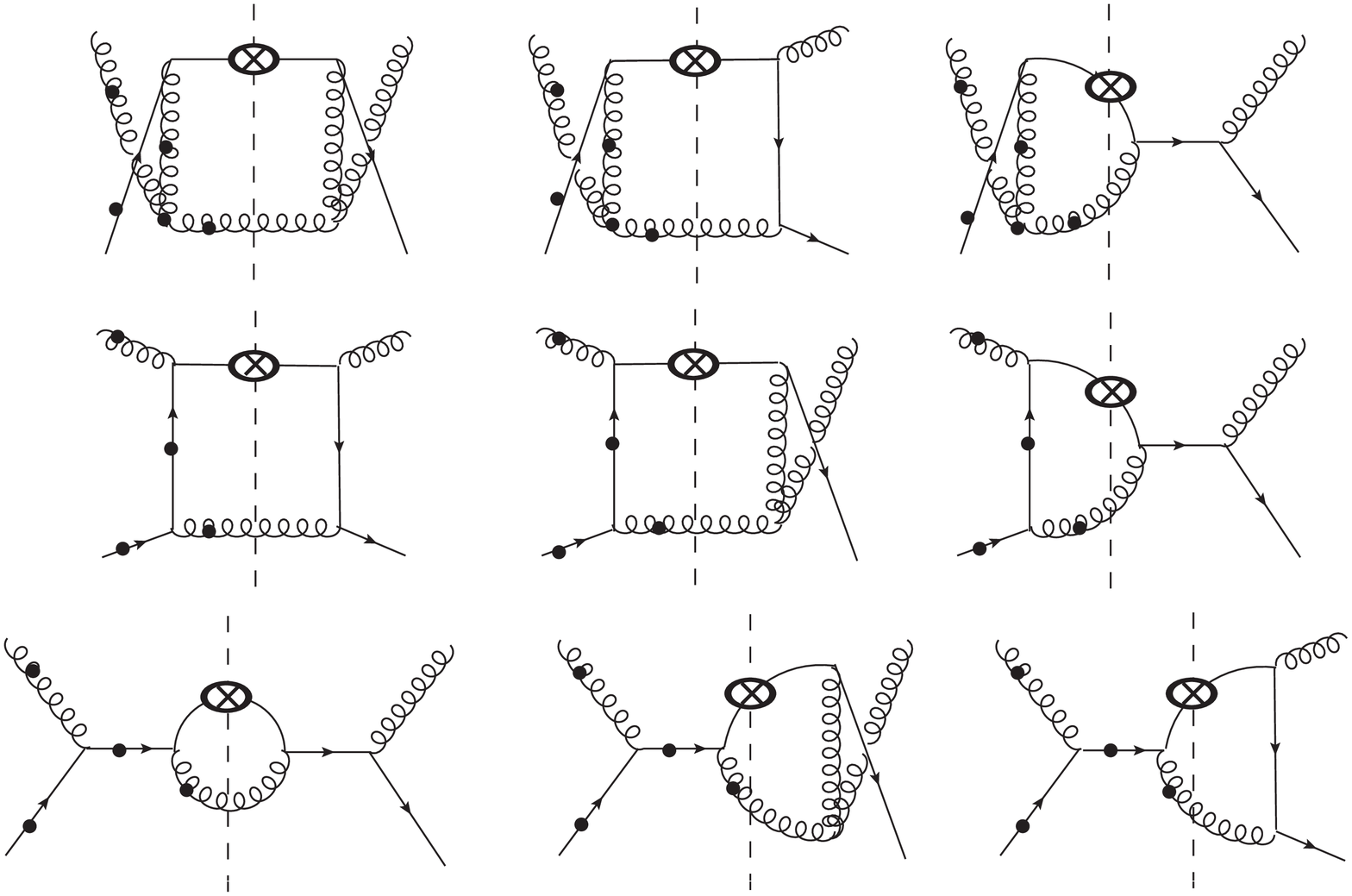}\hspace{1cm}
\end{center}
\vspace{-0.5cm}
\caption{Same as Fig.~\ref{f:qq_qq_1}, but for the $qg\to qg$ channel.}
\label{f:qg_qg}
\end{figure} 

\noindent
(1) $qq'\to qq' \,$:
\begin{eqnarray}
\hat{\sigma}_{1}&=&\frac{\sh(\th^2-2\uh^2)}{\th^3\uh}-\frac{1}{N^2}
\frac{2(\sh^3+2\sh^2\uh+\uh^3)}{\th^3\uh},\\
\hat{\sigma}_{2}&=&\frac{\sh^2+\uh^2}{\th^2\uh}-\frac{1}{N^2}\frac{(2\th-\uh)(\sh^2+\uh^2)}{\th^3\uh},\\
\hat{\sigma}_{3}&=&\left(1-\frac{1}{N^2}\right)\frac{\sh^2+\uh^2}{\th^3},\\
\hat{\sigma}_{4}&=&-\frac{\sh(\sh^2+\uh^2)}{\th^3\uh}-\frac{1}{N^2}\frac{2(\sh^2+\uh^2)}{\th^2\uh}.
\end{eqnarray}

\noindent
(2) $q'q\to qq' \,$:
\begin{eqnarray}
\hat{\sigma}_{1}&=&\frac{\sh(2\th^2-\uh^2)}{\th\uh^3}+
\frac{1}{N^2}\frac{2(\sh^3+2\sh^2\th+\th^3)}{\th\uh^3},\\
\hat{\sigma}_{2}&=&-\frac{\sh^2+\th^2}{\th\uh^2}-\frac{1}{N^2}\frac{(\th-2\uh)(\sh^2+\th^2)}{\th\uh^3},\\
\hat{\sigma}_{3}&=&-\left(1-\frac{1}{N^2}\right)\frac{\sh^2+\th^2}{\uh^3},\\
\hat{\sigma}_{4}&=&\frac{\sh(\sh^2+\th^2)}{\th\uh^3}-\frac{1}{N^2}\frac{2(\sh^2+\th^2)}{\th\uh^2}.
\end{eqnarray}

\noindent
(3) $qq\to qq \,$:
\begin{eqnarray}
\hat{\sigma}_{1}&=&\frac{\sh(\th^2-2\uh^2)}{\th^3\uh}-
\frac{2(\sh^3+2\sh^2\uh+\uh^3)}
{N^2\th^3\uh}
+\frac{\sh(2\th^2-\uh^2)}{\th\uh^3}\nn\\
&+&
\frac{2(\sh^3+2\sh^2\th+\th^3)}{N^2\th\uh^3}
+\frac{\sh^2(\th-\uh)}{N\th^2\uh^2}-
\frac{2\sh^2(\th-\uh)}{N^3\th^2\uh^2},\nn\\
\\
\hat{\sigma}_{2}&=&\frac{\sh^2+\uh^2}{\th^2\uh}-\frac{1}{N^2}\frac{(2\th-\uh)(\sh^2+\uh^2)}{\th^3\uh}
-\frac{\sh^2+\uh^2}{\th\uh^2}\nn\\
&-&\frac{1}{N^2}\frac{(\th-2\uh)(\sh^2+\th^2)}{\th\uh^3}
+\frac{1}{N^2}\frac{2\sh^2(\th-\uh)}{\th^2\uh^2},\\
\hat{\sigma}_{3}&=&
\left(1-\frac{1}{N^2}\right)\frac{\sh^2+\uh^2}{\th^3}
-\left(1-\frac{1}{N^2}\right)\frac{\sh^2+\th^2}{\uh^3}\nn\\
&+&\left(\frac{1}{N}-\frac{1}{N^3}\right)\frac{\sh^2(\th-\uh)}{\th^2\uh^2},\\
\hat{\sigma}_{4}&=&-\frac{\sh(\sh^2+\uh^2)}{\th^3\uh}-\frac{1}{N^2}\frac{2(\sh^2+\uh^2)}{\th^2\uh}
+\frac{\sh(\sh^2+\th^2)}{\th\uh^3}\nn\\
&-&\frac{1}{N^2}\frac{2(\sh^2+\th^2)}{\th\uh^2}
+\left(\frac{1}{N}+\frac{1}{N^3}\right)\frac{\sh^2(\th-\uh)}{\th^2\uh^2}.
\end{eqnarray}

\noindent
(4) $q\bar{q}\to q'\bar{q}^{\,\prime} \,$:
\begin{eqnarray}
\hat{\sigma}_{1}&=&-\frac{\th^2-2\th\uh-\uh^2}{\sh^2\th}+\frac{1}{N^2}\frac{2(\th^2+\uh^2)}{\sh\th\uh},\\
\hat{\sigma}_{2}&=&-\frac{\th^2+\uh^2}{\sh^2\th}-\frac{1}{N^2}\frac{2(\th^2+\uh^2)}{\sh\th\uh},\\
\hat{\sigma}_{3}&=&\frac{1}{N^2}\frac{2(\th-\uh)}{\sh^2},\\
\hat{\sigma}_{4}&=&-\frac{\th^2+\uh^2}{\sh^2\th}-\frac{1}{N^2}\frac{2(\th^2+\uh^2)}{\sh\th\uh}.
\end{eqnarray}

\noindent
(5) $\bar{q}q\to q'\bar{q}^{\,\prime} \,$:
\begin{eqnarray}
\hat{\sigma}_{1}&=&-\frac{\th^2+2\th\uh+\uh^2}{\sh^2\uh}-\frac{1}{N^2}\frac{2(\th^2+\uh^2)}{\sh\th\uh},\\
\hat{\sigma}_{2}&=&\frac{\th^2+\uh^2}{\sh^2\uh}+\frac{1}{N^2}\frac{2(\th^2+\uh^2)}{\sh\th\uh},\\
\hat{\sigma}_{3}&=&\frac{1}{N^2}\frac{2(\th-\uh)}{\sh^2},\\
\hat{\sigma}_{4}&=&\frac{\th^2+\uh^2}{\sh^2\uh}+\frac{1}{N^2}\frac{2(\th^2+\uh^2)}{\sh\th\uh},
\end{eqnarray}

\noindent
(6) $q\bar{q}^{\,\prime} \to q\bar{q}^{\,\prime} \,$:
\begin{eqnarray}
\hat{\sigma}_{1}&=&\frac{\sh^2-2\sh\uh-\uh^2}{\sh^3}+\frac{1}{N^2}
\frac{2(\sh^3+2\sh\uh^3+\uh^3)}{\th^3\uh},\\
\hat{\sigma}_{2}&=&-\frac{1}{N^2}\frac{(2\sh+\uh)(\sh^2+\uh^2)}{\th^3\uh},\\
\hat{\sigma}_{3}&=&\frac{\sh^2+\uh^2}{\th^3}-\frac{1}{N^2}\frac{\sh^2+\uh^2}{\th^3},\\
\hat{\sigma}_{4}&=&\frac{\sh^2+\uh^2}{\th^3}+\frac{1}{N^2}\frac{2(\sh^2+\uh^2)}{\th^2\uh}.
\end{eqnarray}

\noindent
(7) $\bar{q}^{\,\prime}q\to q\bar{q}^{\,\prime} \,$:
\begin{eqnarray}
\hat{\sigma}_{1}&=&-\frac{\sh^2-2\sh\th-\th^2}{\uh^3}-
\frac{1}{N^2}\frac{2(\sh^3+2\sh\th^3+\th^3)}{\th\uh^3},\\
\hat{\sigma}_{2}&=&\frac{1}{N^2}\frac{(2\sh+\th)(\sh^2+\th^2)}{\th\uh^3},\\
\hat{\sigma}_{3}&=&-\frac{\sh^2+\th^2}{\uh^3}+\frac{1}{N^2}\frac{\sh^2+\th^2}{\uh^3},\\
\hat{\sigma}_{4}&=&-\frac{\sh^2+\th^2}{\uh^3}-\frac{1}{N^2}\frac{2(\sh^2+\th^2)}{\th\uh^2}.
\end{eqnarray}

\noindent
(8) $q\bar{q}\to q\bar{q} \,$:
\begin{eqnarray}
\hat{\sigma}_{1}&=&-\frac{\th^2-2\th\uh-\uh^2}{\sh^2\uh}+
\frac{2(\th^2+\uh^2)}{N^2\sh\th\uh}+
\frac{\sh^2-2\sh\uh-\uh^2}{\sh^3}\nn\\
&+&
\frac{2(\sh^3+2\sh\uh^2+\uh^3)}{N^2\th^3\uh}+
\frac{\uh(\sh-\th)}{N\sh\th^2}-
\frac{3\uh(\sh-\th)}{N^3\sh\th^2},\\
\hat{\sigma}_{2}&=&-\frac{\th^2+\uh^2}{\sh^2\th}-
\frac{2(\th^2+\uh^2)}{N^2\sh\th\uh}-
\frac{(2\sh+\uh)(\sh^2+\uh^2)}{N^2\th^3\uh}\nn\\
&-&\frac{1}{N}\frac{2\uh}{\sh\th}+\frac{1}{N^3}\frac{2\uh}{\th^2},\\
\hat{\sigma}_{3}&=&\frac{1}{N^2}\frac{2(\th-\uh)}{\sh^2}+\left(1-\frac{1}{N^2}\right)
\frac{\sh^2+\uh^2}{\th^3}\nn\\
&-&\frac{1}{N}\frac{\uh^2}{\sh\th^2}+\frac{1}{N^3}\frac{\uh^2}{\sh\th^2},\\
\hat{\sigma}_{4}&=&-\frac{\th^2+\uh^2}{\sh^2\th}-\frac{1}{N^2}
\frac{2(\th^2+\uh^2)}{\sh\th\uh}+\frac{\sh^2+\uh^2}{\th^3}\nn\\
&+&\frac{1}{N^2}\frac{2(\sh^2+\uh^2)}{\th^2\uh}+\left(\frac{1}{N}+\frac{1}{N^3}\right)
\frac{\uh(\sh-\th)}{\sh\th^2}.
\end{eqnarray}

\noindent
(9) $\bar{q}q\to q\bar{q} \,$:
\begin{eqnarray}
\hat{\sigma}_{1}&=&-\frac{\th^2+2\th\uh-\uh^2}{\sh^2\uh}-
\frac{2(\th^2+\uh^2)}{N^2\sh\th\uh}-\frac{\sh^2-2\sh\th-\th^2}{\uh^3}\nn\\
&-&\frac{2(\sh^3+2\sh\th^2+\th^3)}{N^2\th\uh^3}-
\frac{\th(\sh-\uh)}{N\sh\uh^2}+
\frac{3\th(\sh-\uh)}{N^3\sh\uh^2},\\
\hat{\sigma}_2&=&\frac{\th^2+\uh^2}{\sh^2\uh}+\frac{1}{N^2}\frac{2(\th^2+\uh^2)}
{\sh\th\uh}+
\frac{(2\sh+\th)(\sh^2+\th^2)}{N^2\th\uh^3}\nn\\
&+&\frac{1}{N}\frac{2\th}{\sh\uh}-\frac{1}{N^3}\frac{2\th}{\uh^2},\\
\hat{\sigma}_{3}&=&
\frac{2(\th-\uh)}{N^2\sh^2}
-{N^2-1\over N^2}
\frac{\sh^2+\th^2}{\uh^3}
+\frac{1}{N}\frac{\th^2}{\sh\uh^2}-\frac{1}{N^3}\frac{\th^2}{\sh\uh^2},\nn\\
\\
\hat{\sigma}_{4}&=&\frac{\th^2+\uh^2}{\sh^2\uh}+
\frac{1}{N^2}\frac{2(\th^2+\uh^2)}{\sh\th\uh}-\frac{\sh^2+\th^2}{\uh^3}\nn\\
&-&\frac{1}{N^2}\frac{2(\sh^2+\th^2)}{\th\uh^2}-(\frac{1}{N}+\frac{1}{N^3})\frac{\th(\sh-\uh)}{\sh\uh^2}.
\end{eqnarray}

\noindent
(10) $qg\to qg \,$:
\begin{eqnarray}
\hat{\sigma}_{1} &=&-\frac{2\sh^5+3\sh^4\uh-\sh^3\uh^2+
\sh^2\uh^3-3\sh\uh^4-2\uh^5}{\sh\th^3\uh^2} \nn\\
&+&
\frac{\sh^3+2\sh^2\uh-2\sh\uh^2-\uh^3}{N^2\sh\th\uh^2} \nn\\
&+& \frac{\sh^3-\sh^2\uh+\sh\uh^2-\uh^3}{(N^2-1)\th^3\uh},\\
\hat{\sigma}_2 &=& -\frac{\sh(\sh^2+\uh^2)}{\th^2\uh^2}-
\frac{\sh^2+\uh^2}{N^2\sh\th\uh}
-\frac{\sh^3-\sh^2\uh+\sh\uh^2-\uh^3}{(N^2-1)\th^3\uh},\nn\\
\\
\hat{\sigma}_{3} &=& -\frac{(\sh^2+\uh^2)^2}{\sh\th^3\uh}-\frac{1}{N^2}\frac{1}{\sh},\\
\hat{\sigma}_{4} &=&\frac{\sh^5+\sh^3\uh^2-\sh^2\uh^3-\uh^5}{\sh\th^3\uh^2}-
\frac{1}{N^2}\frac{\sh-\uh}{\th\uh}\nn\\
&-&\frac{1}{N^2-1}\frac{\sh^3-\sh^2\uh+\sh\uh^2-\uh^3}{\th^3\uh}.
\end{eqnarray}

\noindent
(11) $gq\to qg \,$:
\begin{eqnarray}
\hat{\sigma}_{1}&=&\frac{2\sh^5+3\sh^4\th-\sh^3\th^2+\sh^2\th^3-3\sh\th^4-2\th^5}{\sh\th^2\uh^3}\nn\\
&-&\frac{\sh^3+2\sh^2\th-2\sh\th^2-\th^3}{N^2\sh\th^2\uh}-
\frac{\sh^3-\sh^2\th+\sh\th^2-\th^3}{(N^2-1)\th\uh^3},\nn\\
\\
\hat{\sigma}_2&=&\frac{\sh(\sh^2+\th^2)}{\th^2\uh^2}+\frac{1}{N^2}\frac{\sh^2+\th^2}{\sh\th\uh}\nn\\
&+&\frac{1}{N^2-1}\frac{\sh^3-\sh^2\th+\sh\th^2-\th^3}{\th\uh^3},\\
\hat{\sigma}_{3}&=&\frac{(\sh^2+\th^2)^2}{\sh\th\uh^3}+\frac{1}{N^2}\frac{1}{\sh},\\
\hat{\sigma}_{4}&=&-\frac{\sh^5+\sh^3\th^2-\sh^2\th^3-\th^5}{\sh\th^2\uh^3}
+\frac{1}{N^2}\frac{\sh-\th}{\th\uh}\nn\\
&+&\frac{1}{N^2-1}\frac{\sh^3-\sh^2\th+\sh\th^2-\th^3}{\th\uh^3}.
\end{eqnarray}

\noindent
(12) $gg\to q\bar{q}$ :
\begin{eqnarray}
\hat{\sigma}_{1}&=&-\frac{N}{N^2-1}
\frac{(\th-\uh)(2\th^4+5\th^3\uh+4\th^2\uh^2+5\th\uh^3+2\uh^4)}{\sh^2\th^2\uh^2}\nn\\
&+&\frac{1}{N(N^2-1)}\frac{\th^3+2\th^2\uh-\th\uh^2-\uh^3}{\th^2\uh^2}\nn\\
&+&\frac{N}{(N^2-1)^2}\frac{\th^3-\th^2\uh+2\th\uh^2-\uh^3}{\sh^2\th\uh},\\
\hat{\sigma}_2&=&\frac{N}{N^2-1}\frac{(\th^2+\uh^2)(\th^3-\uh^3)}{\sh^2\th^2\uh^2}\nn\\
&-&\frac{N}{(N^2-1)^2}\frac{\th^3-\th^2\uh+\th\uh^2-\uh^3}{\sh^2\th\uh},\\
\hat{\sigma}_{3}&=&-\frac{1}{N(N^2-1)}\frac{\th-\uh}{\th\uh},\\
\hat{\sigma}_{4}&=&\frac{N}{N^2-1}\frac{\th^5+\th^3\uh^2-\th^2\uh^3-\uh^5}{\sh\th^2\uh^2}
-\frac{1}{N^2}\frac{\th-\uh}{\th\uh}\nn\\
&-&\frac{1}{N^2-1}\frac{\th^3-\th^2\uh+\th\uh^2-\uh^3}{\sh^2\th\uh}.
\end{eqnarray}
Note that for a LO calculation of the fragmentation term, more channels contribute to $A_N$ for 
$p \, p \to \Lambda^{\uparrow} X$ than for $p^{\uparrow} p \to \pi X$~\cite{Metz:2012ct} 
due to the chiral-odd nature of the parton correlators involved in the latter case.
\begin{figure}[t]
\begin{center}
\includegraphics[width=7cm]{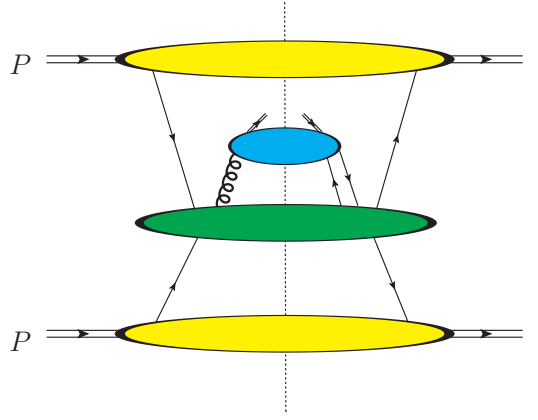}
\end{center}
\vspace{-0.5cm}
\caption{Additional twist-3 fragmentation contribution to $p \, p\to\Lambda^{\uparrow}X$ 
which is not included in the present study.}
\label{f:t3ff_g}
\end{figure} 

\section{Summary}
\label{s:summary}
We have calculated, to leading order in perturbation theory, the twist-3 fragmentation contribution to the 
transverse SSA $A_N$ for hyperon production in unpolarized proton-proton collisions.
Specifically, we have taken into account all contributions arising from {\it qq} and {\it qgq} fragmentation correlators.
We have verified that the result of the cross section is frame-independent when taking into account relations 
between the twist-3 FFs which are based on the QCD equation of motion and the Lorentz invariance of the parton 
correlators.

In order to complete the calculation of this spin-dependent twist-3 fragmentation effect, one also needs to 
include $q\bar{q}g$ correlators (see Fig.~\ref{f:t3ff_g}) as well as {\it gg} and {\it ggg} correlators.  
We mention that the $q\bar{q}g$ graphs need the pure gluon ones in order to have a gauge invariant subset of diagrams.
(In the case of $A_N$ for $p^{\uparrow} p \to \pi X$ the contribution from the former vanishes after summing 
over all graphs, while the latter do not contribute at all~\cite{Metz:2012ct}.)
Let us finally mention that so far the properties of tri-gluon FFs have only been studied to some extent~\cite{Meissner:thesis}.
In particular, the LIRs in this case are not yet known.
We plan to address these topics elsewhere.

\begin{acknowledgments}
This work has been supported by the Grant-in-Aid for
Scientific Research from the Japanese Society of Promotion of Science
under Contract No.~26287040 (Y.K.), 
the National Science Foundation under Contract No.~PHY-1516088 (A.M.),
the U.S. Department of Energy, Office of Science, Office of Nuclear Physics within 
the framework of the TMD Topical Collaboration (D.P.), 
and in part by the U.S. Department of Energy, Office of Science under contract 
No. DE-AC52-06NA25396 and the LANL LDRD Program (S.Y.).
\end{acknowledgments}



\begin{thebibliography}{99}
\bibitem{Bunce:1976yb} 
  G.~Bunce {\it et al.},
  Phys.\ Rev.\ Lett.\  {\bf 36}, 1113 (1976).

\bibitem{Heller:1978ty} 
  K.~J.~Heller {\it et al.},
  Phys.\ Rev.\ Lett.\  {\bf 41}, 607 (1978)
  Erratum: Phys.\ Rev.\ Lett.\  {\bf 45}, 1043 (1980).

\bibitem{Erhan:1979xm} 
  S.~Erhan {\it et al.},
  Phys.\ Lett.\  {\bf 82B}, 301 (1979).

\bibitem{Heller:1983ia} 
  K.~J.~Heller {\it et al.},
  Phys.\ Rev.\ Lett.\  {\bf 51}, 2025 (1983).

\bibitem{Lundberg:1989hw} 
  B.~Lundberg {\it et al.},
  Phys.\ Rev.\ D {\bf 40}, 3557 (1989).

\bibitem{Yuldashev:1990az} 
  B.~S.~Yuldashev {\it et al.},
  Phys.\ Rev.\ D {\bf 43}, 2792 (1991).

\bibitem{Ramberg:1994tk} 
  E.~J.~Ramberg {\it et al.},
  Phys.\ Lett.\ B {\bf 338}, 403 (1994).

\bibitem{Fanti:1998px} 
  V.~Fanti {\it et al.},
  Eur.\ Phys.\ J.\ C {\bf 6}, 265 (1999).

\bibitem{Abt:2006da} 
  I.~Abt {\it et al.} [HERA-B Collaboration],
  Phys.\ Lett.\ B {\bf 638}, 415 (2006)
  [hep-ex/0603047].

\bibitem{Aaij:2013oxa} 
  R.~Aaij {\it et al.} [LHCb Collaboration],
  Phys.\ Lett.\ B {\bf 724}, 27 (2013)
  [arXiv:1302.5578 [hep-ex]].

\bibitem{ATLAS:2014ona} 
  G.~Aad {\it et al.} [ATLAS Collaboration],
  Phys.\ Rev.\ D {\bf 91}, 032004 (2015)
  [arXiv:1412.1692 [hep-ex]].

\bibitem{Pondrom:1985aw} 
  L.~G.~Pondrom,
  Phys.\ Rept.\  {\bf 122}, 57 (1985).

\bibitem{Panagiotou:1989sv} 
  A.~D.~Panagiotou,
  Int.\ J.\ Mod.\ Phys.\ A {\bf 5}, 1197 (1990).

\bibitem{Metz:2016swz} 
  A.~Metz and A.~Vossen,
  Prog.\ Part.\ Nucl.\ Phys.\  {\bf 91}, 136 (2016)
  [arXiv:1607.02521 [hep-ex]].

\bibitem{Aston:1981em} 
  D.~Aston {\it et al.} 
  Nucl.\ Phys.\ B {\bf 195}, 189 (1982).

\bibitem{Abe:1983jy} 
  K.~Abe {\it et al.} 
  Phys.\ Rev.\ D {\bf 29}, 1877 (1984).

\bibitem{Airapetian:2007mx} 
  A.~Airapetian {\it et al.} [HERMES Collaboration],
  Phys.\ Rev.\ D {\bf 76}, 092008 (2007)
  [arXiv:0704.3133 [hep-ex]].

\bibitem{Airapetian:2014tyc} 
  A.~Airapetian {\it et al.} [HERMES Collaboration],
  Phys.\ Rev.\ D {\bf 90}, 072007 (2014)
  [arXiv:1406.3236 [hep-ex]].

\bibitem{Ackerstaff:1997nh} 
  K.~Ackerstaff {\it et al.} [OPAL Collaboration],
  Eur.\ Phys.\ J.\ C {\bf 2}, 49 (1998)
  [hep-ex/9708027].

\bibitem{Abdesselam:2016nym} 
  A.~Abdesselam {\it et al.} [Belle Collaboration],
  arXiv:1611.06648 [hep-ex].

\bibitem{Kane:1978nd} 
  G.~L.~Kane, J.~Pumplin and W.~Repko,
  Phys.\ Rev.\ Lett.\  {\bf 41}, 1689 (1978).

\bibitem{Efremov:1981sh} 
  A.~V.~Efremov and O.~V.~Teryaev,
  Sov.\ J.\ Nucl.\ Phys.\  {\bf 36}, 140 (1982)
  [Yad.\ Fiz.\  {\bf 36}, 242 (1982)].

\bibitem{Ellis:1982wd} 
  R.~K.~Ellis, W.~Furmanski and R.~Petronzio,
  Nucl.\ Phys.\ B {\bf 207}, 1 (1982).

\bibitem{Ellis:1982cd} 
  R.~K.~Ellis, W.~Furmanski and R.~Petronzio,
  Nucl.\ Phys.\ B {\bf 212}, 29 (1983).

\bibitem{Efremov:1984ip} 
  A.~V.~Efremov and O.~V.~Teryaev,
  Phys.\ Lett.\  {\bf 150B}, 383 (1985).

\bibitem{Qiu:1991pp} 
  J.~w.~Qiu and G.~F.~Sterman,
  Phys.\ Rev.\ Lett.\  {\bf 67}, 2264 (1991).

\bibitem{Qiu:1991wg} 
  J.~w.~Qiu and G.~F.~Sterman,
  Nucl.\ Phys.\ B {\bf 378}, 52 (1992).

\bibitem{Qiu:1998ia} 
  J.~w.~Qiu and G.~F.~Sterman,
  Phys.\ Rev.\ D {\bf 59}, 014004 (1999)
  [hep-ph/9806356].

\bibitem{Kanazawa:2000kp} 
  Y.~Kanazawa and Y.~Koike,
  Phys.\ Lett.\ B {\bf 490}, 99 (2000)
  [hep-ph/0007272].

\bibitem{Eguchi:2006qz}
  H.~Eguchi, Y.~Koike and K.~Tanaka,
  Nucl.\ Phys.\  B {\bf 752}, (2006)
  [arXiv:hep-ph/0604003].

\bibitem{Eguchi:2006mc} 
  H.~Eguchi, Y.~Koike and K.~Tanaka,
  Nucl.\ Phys.\ B {\bf 763}, 198 (2007)
  [hep-ph/0610314].

\bibitem{Kouvaris:2006zy} 
  C.~Kouvaris, J.~W.~Qiu, W.~Vogelsang and F.~Yuan,
  Phys.\ Rev.\ D {\bf 74}, 114013 (2006)
  [hep-ph/0609238].

\bibitem{Koike:2009ge} 
  Y.~Koike and T.~Tomita,
  Phys.\ Lett.\ B {\bf 675}, 181 (2009)
  [arXiv:0903.1923 [hep-ph]].
  
\bibitem{Metz:2012ct} 
  A.~Metz and D.~Pitonyak,
  Phys.\ Lett.\ B {\bf 723}, 365 (2013)
  Erratum: Phys.\ Lett.\ B {\bf 762}, 549 (2016)
  [arXiv:1212.5037 [hep-ph]].
  
\bibitem{Beppu:2013uda} 
  H.~Beppu, K.~Kanazawa, Y.~Koike and S.~Yoshida,
  Phys.\ Rev.\ D {\bf 89}, 034029 (2014)
  [arXiv:1312.6862 [hep-ph]].
  
\bibitem{Pitonyak:2016hqh} 
  D.~Pitonyak,
  Int.\ J.\ Mod.\ Phys.\ A {\bf 31}, 1630049 (2016)
  [arXiv:1608.05353 [hep-ph]].
    
 \bibitem{Kanazawa:2000cx} 
  Y.~Kanazawa and Y.~Koike,
  Phys.\ Rev.\ D {\bf 64}, 034019 (2001)
  [hep-ph/0012225].
  
\bibitem{Zhou:2008fb} 
  J.~Zhou, F.~Yuan and Z.~T.~Liang,
  Phys.\ Rev.\ D {\bf 78}, 114008 (2008)
  [arXiv:0808.3629 [hep-ph]].
  
\bibitem{Koike:2015zya} 
  Y.~Koike, K.~Yabe and S.~Yoshida,
  Phys.\ Rev.\ D {\bf 92}, 094011 (2015)
  [arXiv:1509.06830 [hep-ph]].

\bibitem{Yabe:talk}
  K.~Yabe, talk given at the 22nd International Spin Physics Symposium (SPIN 2016).
  
\bibitem{Yabe:proceedings}  
  K.~Yabe, Y.~Koike, A.~Metz, D.~Pitonyak, and S.~Yoshida,
  in Proc.~of the 22nd International Spin Physics Symposium (SPIN 2016). 
 
 
 \bibitem{Kanazawa:2014dca} 
  K.~Kanazawa, Y.~Koike, A.~Metz and D.~Pitonyak,
  Phys.\ Rev.\ D {\bf 89}, 111501(R) (2014)
  [arXiv:1404.1033 [hep-ph]].

\bibitem{Gamberg:2017gle} 
  L.~Gamberg, Z.~B.~Kang, D.~Pitonyak and A.~Prokudin,
  arXiv:1701.09170 [hep-ph].
 
 \bibitem{Kanazawa:2015ajw} 
  K.~Kanazawa, Y.~Koike, A.~Metz, D.~Pitonyak and M.~Schlegel,
  Phys.\ Rev.\ D {\bf 93}, 054024 (2016)
  [arXiv:1512.07233 [hep-ph]].
  
\bibitem{Mulders:1995dh} 
  P.~J.~Mulders and R.~D.~Tangerman,
  Nucl.\ Phys.\ B {\bf 461}, 197 (1996)
  Erratum: Nucl.\ Phys.\ B {\bf 484}, 538 (1997)
  [hep-ph/9510301].

\bibitem{Bacchetta:2006tn} 
  A.~Bacchetta, M.~Diehl, K.~Goeke, A.~Metz, P.~J.~Mulders and M.~Schlegel,
  JHEP {\bf 0702}, 093 (2007)
  [hep-ph/0611265].

\bibitem{Anselmino:2000vs} 
  M.~Anselmino, D.~Boer, U.~D'Alesio and F.~Murgia,
  Phys.\ Rev.\ D {\bf 63}, 054029 (2001)
  [hep-ph/0008186].

\bibitem{Anselmino:2001js} 
  M.~Anselmino, D.~Boer, U.~D'Alesio and F.~Murgia,
  Phys.\ Rev.\ D {\bf 65}, 114014 (2002)
  [hep-ph/0109186].

\bibitem{Astier:2000ax} 
  P.~Astier {\it et al.} [NOMAD Collaboration],
  Nucl.\ Phys.\ B {\bf 588}, 3 (2000).

\bibitem{Meissner:thesis}
  S.~Meissner,
  Ph.D.~Thesis, University of Bochum, 2009.
  
 \bibitem{Kanazawa:2015jxa} 
  K.~Kanazawa, A.~Metz, D.~Pitonyak and M.~Schlegel,
  Phys.\ Lett.\ B {\bf 744}, 385 (2015)
  [arXiv:1503.02003 [hep-ph]].

\bibitem{Gamberg:2008yt} 
  L.~P.~Gamberg, A.~Mukherjee and P.~J.~Mulders,
  Phys.\ Rev.\ D {\bf 77}, 114026 (2008)
  [arXiv:0803.2632 [hep-ph]].

\bibitem{Meissner:2008yf} 
  S.~Meissner and A.~Metz,
  Phys.\ Rev.\ Lett.\  {\bf 102}, 172003 (2009)
  [arXiv:0812.3783 [hep-ph]].

\bibitem{Gamberg:2010uw} 
  L.~P.~Gamberg, A.~Mukherjee and P.~J.~Mulders,
  Phys.\ Rev.\ D {\bf 83}, 071503 (2011)
  [arXiv:1010.4556 [hep-ph]].

\bibitem{Kanazawa:2013uia} 
  K.~Kanazawa and Y.~Koike,
  Phys.\ Rev.\ D {\bf 88}, 074022 (2013)
  [arXiv:1309.1215 [hep-ph]].

\bibitem{Metz:2002iz} 
  A.~Metz,
  Phys.\ Lett.\ B {\bf 549}, 139 (2002)
  [hep-ph/0209054].

\bibitem{Collins:2004nx} 
  J.~C.~Collins and A.~Metz,
  Phys.\ Rev.\ Lett.\  {\bf 93}, 252001 (2004)
  [hep-ph/0408249].

\bibitem{Yuan:2009dw} 
  F.~Yuan and J.~Zhou,
  Phys.\ Rev.\ Lett.\  {\bf 103}, 052001 (2009)
  [arXiv:0903.4680 [hep-ph]].

\bibitem{Kang:2010zzb} 
  Z.~B.~Kang, F.~Yuan and J.~Zhou,
  Phys.\ Lett.\ B {\bf 691}, 243 (2010)
  [arXiv:1002.0399 [hep-ph]].

\bibitem{Kanazawa:2014tda} 
  K.~Kanazawa, A.~Metz, D.~Pitonyak and M.~Schlegel,
  Phys.\ Lett.\ B {\bf 742}, 340 (2015)
  [arXiv:1411.6459 [hep-ph]].

\end{thebibliography}

\end{document}